\documentstyle[aps,prl,preprint,epsfig]{revtex}
\tightenlines
\title{ Vlasov Description Of Dense Quark Matter }
\author{A. Bonasera\footnote{bonasera@lns.infn.it}}
\address{
Laboratorio Nazionale del Sud, Istituto Nazionale di Fisica Nucleare, 
Via S. Sofia 44, 95123 Catania, Italy $\&$\\
Cyclotron Institute, Texas $A\&M$ University, College Station, 
TX 77843-3366,USA.
 }
\begin{document}
\maketitle
\begin{abstract}
We discuss properties of quark matter at finite baryon densities and
zero temperature in a Vlasov approach. 
We use a screened interquark Richardson's potential
consistent with the indications of Lattice QCD calculations.
  We analyze the choices of the quark masses and the
parameters entering the potential which reproduce the binding energy (B.E.) of
infinite nuclear matter. There is a transition from nuclear to quark matter
at densities 5 times above normal nuclear matter density. 
The transition could be revealed from the determination of the
 position of the shifted meson masses in dense baryonic matter. A scaling
form of the meson masses in dense matter is given.
\end{abstract}

%\newpage

{\ \vskip 2\baselineskip

% body of the paper
\newpage

%\narrowtext
%\begin{multicols}{2}
\section{Introduction}

The search for a quark-gluon plasma (QGP) is one of the new and most exciting
directions in physics at the border between nuclear and particle physics.
One way to get to the QGP is via the collisions of relativistic heavy ions
and this will be accomplished soon at the Relativistic Heavy Ion Collider
(RHIC) at Brookhaven and in the next decade at LHC-CERN \cite{wong}.
This strategy, the only possible on earth, raises many problems and 
difficulties which could be resolved with the help of theory.  Some
of the difficulties are:
i) the collision is mainly a non-equilibrium process.  Equilibration-even
 partial- might not be reached during the process and the nuclei could
be transparent at very high energies;
ii) the properties of the particles change when the density and/or the 
excitation energy (temperature?) change; iii) it is not
clear which 'probes' will give relevant informations on the dynamics
of the collision.
Thus a theoretical approach must be dynamical and contain the
most important features of microscopic approaches like Quantum ChromoDynamics
(QCD) which because of its difficulties (numerical and conceptual),
has been applied so far to some
limited cases such as quark matter at zero baryon density and high temperature
(T) \cite{wong,pov}.  Recently \cite{bon99}, 
we have proposed a dynamical approach based on the
Vlasov equation \cite{repo,land1} to reproduce hadron masses. 
Some work in the same spirit is  discussed in \cite{mosel}.
The approach
needs as inputs the interaction potential among quarks which we borrowed
from phenomenology i.e. the Richardson's potential\cite{rich} and the quark 
masses which we fitted to reproduce 
known meson masses such as the $\pi$, the $\phi$, the 
$\eta_c$ etc.  When the particles are embedded in a dense medium such as
in nuclear matter (NM)the potential becomes screened in a similar fashion
as ions and electrons in condensed matter do, i.e. 
Debye screening (DS)\cite{wong,land1}. 
Thus we use a screened potential
and search for the relevant parameters which reproduce
the properties of NM i.e. a minimum at
baryon density $\rho_0=0.16fm^{-3}$ and an energy per nucleon 
$E/nucleon=-16.MeV$.    Then 
we can study the equation of state (EOS) of quark matter at various baryon
densities ($\rho$) 
and determine the transition from NM to QGP.  All the calculations
in this paper are at finite $\rho$ and zero T,
but an extension to other cases and to heavy ion collisions is possible 
and will  be discussed in future works.  In brief these are our main results:

a) the approach can reproduce the well known properties of NM near the
ground state density.  This has the important consequence that 
we can easily extend the 
calculations to finite nuclei starting from the quark degrees of freedom;

b) the transition between NM and QGP is smooth, we find
no evidence for a first order phase transition. Of course
the results could change at finite T;

c) some features of QM can be revealed by studying the properties
of the hadrons in dense medium.   Mesons such as
the $\pi$, $J/\psi$ and $\gamma(1S)$ can survive in the medium up
to a "critical" density which depends on the particles type.  We also
find that light quark mesons such as the $\rho$ $\it never$ survive in the
medium because of the repulsive chromomagnetic term(c.t.), see eq.(3) below. 
 For those particles
that can be bound in QM we can express their masses m($\rho$) in medium as:
\begin{equation}
m_c-m(\rho)\propto (1-\frac{\rho}{\rho_c})^{\beta}
\end{equation}
where $m_c$ and $\rho_c$ are the values of the mass and baryon density
where the mesons dissolve in medium and $\beta$ is a fitting parameter.
We stress that eq.(1) has nothing to do with the behavior of an order parameter
near the critical point in a second order phase transition.
  A scaling of the
hadron masses in a dense medium has been shown by Brown and Rho
\cite{brown} based on an effective chiral Lagrangian and the scaling
properties of QCD. Also the study of the properties of meson masses in a
medium, especially the $J/\psi$ has been discussed to signal the QGP
phase transition at high T\cite{satz}.

The paper is organized as follows.  In section II we review the Vlasov
 equation and the test particles method.  In section III we apply the
method to calculate hadron masses while section IV is devoted to
the EOS calculations.  A brief summary is given in section V.
\section{The Vlasov Equation}
   We briefly recall some general features of the VE, 
interested  people should look the (not complete) list of
    references for more details on this equation
   \cite{wong,bon99,repo,land1,mosel}.  
The VE gives the time evolution of the one
   body distribution function $f_{q_c}(r,p,t)$ in phase-space:
   
   \begin{equation}
\partial_{t}f_{q_c}+\frac{\overrightarrow{\text{p}}}{E}\cdot \nabla _{r}f_{q_c}
-\nabla
_{\text{r}}U\cdot \nabla _{\text{p}}f_{q_c}=0 
\label{lv}
\end{equation}

where 
$E=\sqrt{p^2+m_i^2}$ is the energy and 
$m_i$ is the quark mass. The underscripts  indicate
that the distribution function
 depends on the flavor (q) and color (c) of the quarks.    
 
In agreement to LQCD calculations\cite{pov,rich} the interacting potential
$U(r)$ for $q\bar q$ is ($\hbar=1$):
 
 \begin{eqnarray}
 U(r)=\frac{8\pi}{33-2n_f}\Lambda(\Lambda r-\frac{f(\Lambda r)}{\Lambda r})
+\frac{8\pi}{9}
\bar{\alpha}\frac{<{\bf \sigma}_q{\bf \sigma}_{\bar q}>}
{m_q m_{\bar q}}\delta({\bf r}) 
 \end{eqnarray}
 
 where 
 
 \begin{eqnarray}
f(t)=1-4 \int{\frac{dq}{q}\frac{e^{-qt}}{[ln(q^2-1)]^2+\pi^2}} 
 \end{eqnarray}

( for $qq$ the potential must be decreased of a factor 1/2),
$n_f$ is the number of quark flavors involved and
the parameter $\Lambda$ has been fixed to reproduce the masses of heavy 
$c \bar c$ and $b \bar b$ systems in \cite{rich}.
In eq.(3) we have added to the Richardson's potential 
the c.t., very important
 to explain the masses of
different resonances for light quarks in vacuum.  
In this work the expectation value of $<{\bf\sigma}_q
{\bf\sigma}_{\bar q}>$ is used depending on the relative spin orientations of
the constituent quarks . For instance for the pion this term is
equal to -3 while for a $\rho$ meson it is equal to +1\cite{pov}.  The
$\delta$ function is approximated to a gaussian i.e. we make the 
replacement
$\delta({\bf r})\rightarrow \frac{1}{(2\pi\sigma^2)^{3/2}}e^{-\frac{r^2}
{2 \sigma^2}}$.  The average strong coupling constant $\bar \alpha$ and
the variance 
$\sigma$ are adjusted to reproduce the $\pi$ and $\rho$ mesons
\cite{bon99}.  In dense nuclear matter the c.t. can be neglected.

Numerically the VE equation is solved by writing the one body distribution
function as:

\begin{equation}
\label{efer}
f_{q_c}(r,p,t) = \frac{1}{n_{tp}} \sum_i^N g_r(r-r_i(t)) g_p(p-p_i(t)) 
\end{equation}

where the $g_r$ and $g_p$ are sharply peaked distributions (such
 as delta functions, gaussian or other simple
functions), that we shall treat as delta functions.
 $N=Qn_{tp}$ is the number of such terms, $Q={q}+\bar q$ is
the total number of quarks and antiquarks (for a meson Q=2).  Actually, 
N is much larger than the total quark
number Q, so that we can say that each quark 
is represented by ${n_{TP}}$ terms called test particles(tp).
The rigorous mean field limit
can be obtained for ${n_{TP} \rightarrow \infty}$ where the calculations are
of course numerically
 impossible, even though the numerical results converge rather quickly.
 Inserting eq.(~\ref{efer}) in the Vlasov equation
 gives
the Hamilton equations of motions for the tp~\cite{repo}:

\begin{eqnarray}
\dot {\bf r}_i = \frac{{\bf p}_i}{E_{i}} \nonumber \\
\dot {\bf p}_i  = - {{\bf\nabla}_{r_i}} U ,   \label{ripi}
\end{eqnarray}
 
 for $ i=1....N$. The total number of tp used in this work
ranges from 5000 to 50000 with no appreciable change in the results.
The equations of motion (6) 
are solved by using a O($\delta t^4 $) Adams-Bashfort
method ~\cite{koon}.

\section{Hadron Masses}
We recall some results obtained in \cite{bon99}.  In order to calculate
hadron masses, we initially distribute 
  randomly  the tp in a sphere of radius r in coordinate
space and $p_f$ in momentum space.  $p_f$ is the Fermi momentum
estimated in a simple Fermi gas model by imposing that a cell in
phase space of size $h=2 \pi$ can accommodate at most two identical quarks
 of different spins. A
simple estimate gives the following relation between the quarks density
$n_q$ and the Fermi momentum:
\begin{eqnarray}
n_q=\frac{g_q}{6\pi^2}p_f^3 
 \end{eqnarray}
a similar formula holds for $\bar q$.  The degeneracy number 
$g_q=n_c\times n_s \times n_f$, where $n_c$ is the number of colors and
$n_s$ is the number of spins\cite{wong}.  For quarks and
 antiquarks 3 different colors are used red,green and blue (r,g,b) \cite{pov}.  
  From the above equation
we see that the Fermi momentum for quarks distributed in a sphere of radius
0.5 fm is of the order of 0.5 GeV/c. Thus relativistic effects become important
for quark masses less than 1 GeV. For instance for the $\pi$ case
, if we calculate the total energy of the system 
relativistically we obtain 0.14 GeV (with the parameters choice discussed
 below),
 while using the nonrelativistic limit we obtain about 1 GeV!  
We stress that since the system is properly antisymmetrized at time t=0fm/c, 
it will remain so at all times since the VE conserves the volume in phase-space
\cite{repo}.

The masses of the hadrons are given by the total energy of the system. 
It is the interplay among the Fermi motion, 
the potential term and the quark masses 
which determines the masses of the hadrons.   
For light quarks the chromomagnetic term is important
to reproduce the experimental values of the masses.  
Thus we have adjusted the values
of the scale constant and the 
quark masses to reproduce the data.  We found a good fit for 
mesons by using $\Lambda=0.250  GeV$,
 $\bar \alpha_s=0.225$, $\sigma=0.5 fm$ \cite{pov} and
the quark mass values given in Table I.  We notice that
the parameters and the heavy quark mass values 
are in good agreement with  currently accepted ones \cite{pov,data}. The masses
of (u,d) quarks in table I is somewhat smaller than the 300 MeV 
used in many potential models \cite{pov}.  This is due to the fact that these
models are non relativistic but, as we stressed above, because of the Fermi
motion, relativistic effects are quite important.
In the relativistic approach of \cite{crat}, where the Richardson's potential
was used as well, the (u,d,s) quark masses have values comparable to ours. 
The small differences 
between our results and \cite{crat} are most probably due to their
relativistic generalization of the Richardson's potential and the
different choice of $\Lambda$.

 The parameters entering our model are essentially fixed on some
meson masses.  As a consistency check we extended the calculations
to the baryons case with the usual modifications of a factor $1/2$ 
to the potential eq.(4), as 
suggested by LQCD considerations\cite{pov}.  

In figure (1) we display the calculated (open symbols)
mass of the resonances vs. the sum of the quark 
masses (cfr. table I),
for mesons (top) and baryons (bottom).  The circle symbols refer to 
attractive, while the squares to repulsive chromomagnetic term in eq.(4)
\cite{pov}. The corresponding 
experimental data \cite{data} are given by the full symbols.
  The overall agreement
is quite good in all cases and some predictions for resonances not yet 
observed are also given.

The results discussed above prove that the VE is suitable to
describe the quarks dynamics in hadrons despite the simple form for
the two body potential.  The approach works rather well for heavy quarks, in
that it is able to reproduce the masses and radii of the heavy hadrons
(as compared to data or other calculations \cite{pov,bon99,rich}) and also the
 masses of light hadrons.  The radii of light quarks systems are
 underestimated  which could be a hint for relativistic corrections to the 
potential term as discussed for instance in \cite{crat}.
However some criticism can be raised such as:

i) it is commonly accepted that the main features of light quarks are
 dictated by
chiral dynamics.  This is of course  a limit of the Vlasov approach where
a non-relativistic potential is employed.
However, we have shown that a good choice of the parameters can still reproduce
the experimental data.  One should keep in mind that the purpose of this
work is to describe {\it average} properties that could be observed in nucleus
-nucleus collisions and to give some guidance to unveil the occurrence of
a phase transition (if any).  Thus we need a model that gives similar
masses when the particles are isolated,see fig.1,
 plus the confining properties.
Confinenement arises here because of the linear term in the Richardson's
potential, eq.(3).  Furthermore, we require that dense quark matter has
similar properties in our approach as nuclear matter near the ground state
 baryon density, see next section.  Since we will be looking for general 
features such as phase transitions, these features will be similar because
the EOS and the rest masses of hadrons are similar;

ii) in the Vlasov approach color neutral states are the result of the
initial conditions, i.e. at time t=0 fm/c the system is colorless.  The time
evolution will keep the system colorless in each region because
of the large number of tp. The color degrees of freedom enters only
through the Fermi momentum, eq.(7) and they are not contained in the
 potential term.  The only consequence is that if initially one gives
1 or 2 colors for instance to baryons, the Fermi momenta will be higher
and so will be the masses;  

iii) the initial conditions are given by a Fermi gas.  This will result
in a total energy (which is constant) for a given initial radius.  Of
course since the system is finite it will evolve and develop a smooth
surface.  For long enough times the density distribution will evolve
to an equilibrium one, see fig.(1) in \cite{bon99}.  High momentum
states of particles arise when two tp are very close and see a strong
attraction due to the coulomb term in the potential.

\section{Dense Matter}

The potential becomes screened in dense matter because of the
interactions of the many colored quarks  which make a colorless object.  
We incorporate this feature simply by "screening" the Richardson's potential as:
\begin{equation}
U(r)\rightarrow U(r)e^{-\frac{r}{r_d}}
\end{equation}
where $r_d$ is the Debye radius given in dense matter at zero temperature
by\cite{bombaci}:  
\begin{equation}
(r_d)^{-2}=\frac{2\alpha_0}{\pi}\sum_{i=u,d,s..} p_{f_i} \sqrt{p_{f_i}^2+m_i^2}
\end{equation}

where $\alpha_0$ is the strong coupling constant and $p_{f_i}$ is the Fermi
momentum given above, eq.7.  The strong coupling constant is in general
dependent on the relative distances or momenta of the quarks.  
Here, we treat $\alpha_0$ as a constant and fix its
value to reproduce the BE of nuclear matter at $\rho_0$.
We will perform calculations for $Q=300$
 (u-d) quarks and 30 tp per quark,
increasing these numbers does not change the results.
   The initial conditions are given
 by  randomly distributing the tp in a cube of side L in coordinate
space (with periodic boundary conditions) and a sphere of radius 
$p_{f_i}$ in momentum space,  
  we fix $n_f=3$\cite{bombaci}.

In figure(2) we display the calculated energy per nucleon (1 nucleon=3 quarks)
(top part), Debye radius (middle) and
energy density of the quarks $\epsilon$(bottom) vs. $\rho/\rho_0$.  
The $\alpha_0$
has been fitted to give the correct BE at $\rho_0$ for each case.
%From the energy density we can easily calculate
%the energy per baryon $E/nucleon=\epsilon/\rho-939 (MeV)$.  
The open symbols
refer to the choices $\Lambda=100MeV$,
 $m_u=m_d=7 MeV$ (circles) and $m_u=m_d=130 MeV$ (squares).  
For these cases the
minimum in the E/nucleon 
occurs at about twice $\rho_0$ and deeper
BE. The corresponding $r_D$ are of the order of 1 fm at $\rho_0$ which is 
quite a large value.  Recall that at such densities the distance among
{\it nucleons} is also of the order of 1 fm and we would expect $r_D$
to be smaller.  Thus we can exclude these sets of parameters from the
following considerations.  The correct properties of NM at $\rho_0$ 
can be obtained for
instance for the choice $\Lambda=250MeV$,$m_u=m_d=130 MeV$, 
$\alpha_0=1.43$(triangles-set I)
and $\Lambda=400 MeV$, $m_u=m_d=7 MeV$, $\alpha_0=2.26$(diamonds-set II)
\cite{nota}.  Notice that set I has been employed in section III to reproduce
the hadrons masses.  
The E/nucleon can be estimated by using a Skyrme potential with compressibility 
$K=225$ $MeV$ (dashed line
in figure 2 (top))which gives a good description of NM near the gs density
\cite{wong,repo}.
  We notice a deviation from the Skyrme result at high densities 
where the calculated 
EOS becomes softer because of the transition to the QGP.  Another 
deviation occurs at very low densities which hints to a larger
screening in order to reduce the effect of the linear term in the
Richardson's potential.  However both limits are in the regions where our
knowledge of the EOS is rather rough and the Skyrme approximation is 
questionable.
  The corresponding $r_D$ 
and energy densities (in units of the energy densities obtained
in a Fermi gas model $\epsilon_{FG}=\frac{g_q}{8\pi^2}p_{f_i}^4$\cite{wong})
are displayed in the middle and bottom part of the figure. At $\rho_0$ the
$r_D$ are about 0.5 fm and the energy density is of course correct.
 The calculated energy densities cross 
the Fermi gas result at about 5-7 times normal
nuclear matter and tend to the Fermi gas value from below at very high 
densities because the attractive part of the potential is dominant there. 
  The $\epsilon$ is a smooth function of $\rho$
(and $p_f$,eq.7) thus excluding a first order phase  transition.
 Since this plot gives no evidence for a "critical" change of phase,
we will loosely speak of nuclear matter for densities close to the nuclear
matter g.s. density and quark matter at high densities.  A change from one 
phase to another can be estimated from the Bag-model\cite{wong}.
This change will occur when the pressure from the nuclear matter
becomes equal to the bag pressure, i.e. $P_q=B$.  At high densities we can
neglect the quark masses, thus the "critical energy density" 
$\epsilon_c=3 P_q$ \cite{wong}.  Using a value
of $B^{1/4}=0.206 GeV$, gives $\epsilon_c=0.71 GeV$.  This value of energy
density is obtained in our calculations at about $\rho_{cb}=0.75 fm^{-3}$ for
parameters I and II, and it is indicated by the arrow in fig.(2) top.

In order to strenghten this finding we simulate
some 'probes' which behave differently depending on the density. 
  If we put a hadron, in particular a
meson, in NM, its mass will change because of the DS, eq.(8)\cite{brown,satz}.
  In 
particular we can distribute $q\bar q$ in a sphere of radius $r_i$ at time
$t=0 fm/c$ and let it evolve in NM with the screened potential\cite{nota0}.   
In figure(3) we show the total energy of the meson
(i.e. its mass) vs. the initial $r_i$  at different densities 
\cite{wong,satz}, for the $\pi$ (full circles), the 
$\eta_c(1S)$
(triangles) at $\rho_0$ and the heavier $\gamma(1S)$ at $\rho=1.046 fm^{-3}$
(set II).
 Notice that the minima occur at smaller radii for heavier quarks which thus 
probe different densities of the matter. In particular we find different
'critical' densities $\rho_c$ 
for the 3 ($u \bar d ,c \bar c$ and $b \bar b$) cases where
the mesons dissolve.    Another important 
feature is that for light quarks the minima are due to the c.t.,
 infact if we calculate the mass of a $\rho$ meson in medium (for which 
the c.t. is repulsive),
we find no minima at all densities.  In figure (3) we display the results 
for the $\rho$ meson (open circles) 
at $\rho=0.0235 fm^{-3}$.  The difference between the
mass in medium and the free mass is always positive for the cases where
a minimum exists.  This difference increases for increasing densities up to
the $\rho_c$ where the minimum disappears.  
 We stress that for the light $u \bar d$
mesons, the calculated radii in vacuum 
are smaller than data\cite{bon99}, thus
in reality the interplay between the Coulombic, the linear and chromomagnetic 
terms in eq.(3) is different and could change our results, but the qualitative
features should remain unchanged.

The results discussed above are summarized in figure (4) in a scaling 
invariant way.  In the figure we show that the differences of the
'critical' mass $m_c$ 
minus the mass at density $\rho$ vs.
$\rho/\rho_c$ fall in the same universal curve which can be parametrized 
according to eq.(1) and it is given by the full curve in the figure
\cite{nota1}.  It is important to notice that our estimate for $\rho_{cb}$ 
 using the bag model is in good agreement to the critical density obtained
when the $c\bar c$ dissolves in the plasma.  This particle could give important
informations about the density where the QGP is first formed at variance with
lighter quark mesons for which the c.t. might be important and for heavier
resonances which because of the heavy quark mass are less influenced by
 the Debye screening (thus dissolve later).  It is important
to notice however that a direct signature of $c \bar c$ suppression in RHIC
might be washed out by the many interactions that the particles might have
during the reaction time.

In ending this section we would like to stress the following points:

i) we have calculated the EOS starting from a phenomenological q-q potential
 which becomes screened in dense matter.  Of course if the density goes
to zero the screening vanishes and we obtain the confining properties
of the potential (due to the linear term).  This will be important when
dealing with finite nuclei in that it will not be possible to find
isolated quarks, but rather quarks must be always grouped in order to have
sizable densities.   Again the cluster of quarks will be colorless on average
 because of the large number of t.p.  Since this is a mean field approach,
plus we have made some simplifications for the potential, we should
expect the model to work in a reasonable way for average properties such
as collective flow and particles production after a suitable collision term
is added to the Vlasov equation.  Of course we cannot demand that light
nucleon clusters and even nucleons will be formed with the correct quarks
 content.  This is in many ways analogue to the situation in nuclear 
physics where one describes HIC around the Fermi energy using the Boltzmann
 Nordheim Vlasov equation \cite{repo}: average properties are reasonably
reproduced such as collective flow, pion, photon, kaon..production,  but
to reproduce deuteron, triton, and other light heavy ions yields is clearly
 outside the reach of the model;  
 
ii)  in principle one should be able to obtain self-consistently DS  
in the Vlasov approach. 
  However, if we give a different color charge to the t.p.
in such a way that the total color charge is zero and make the bare
 Richardson's potential become repulsive for two equal charges then,
due to the large number of t.p. which results in a smooth quark density 
in the box,  the resulting average potential will be zero.  This could
be corrected by using a different way to initialize the system 
(which we have been unable to do so far) or use a molecular dynamics
approach.  There could be however some problems in molecular dynamics as
well, such as numerical (the CPU time increases quadratically with the number
of particles- at variance with the VE which is linearly dependent).  Most
important it will be very difficult to keep antisymmetrization at all
times since molecular dynamics will evolve the system to its classical 
equilibrium state;

iii)  one could also wonder if in our matter calculations, drops describing
a confined nucleon, say at g.s. density, are really  formed.  This is
not easy to see in the calculations since there is a time evolution and 
blobs of matter are created and destroyed at all times and only {\it 
on average} 
(over time) the density is constant.  However, since the EOS we obtain 
is the same
i.e. same density, binding energy and (by opportunely tuning the
parameters) same compressibility of NM, such a problem does not apply.    
\section{Conclusions}

In conclusion in this work we have calculated the EOS of quark matter within
a Vlasov approach at finite baryon densities and zero temperatures.  We
have shown that a suitable choice of the parameters and quark masses reproduces
the nuclear matter properties near the ground state.  One set of 
parameters (I) used is also able to reproduce hadrons masses
 as well.   
At large densities the
EOS becomes softer than as given by a Skyrme force because of the 
transition to the QGP.  The transition can be estimated comparing our results
 to the bag pressure and energy density.
 The density of the transition is close to the point
where the $c\bar c$ mesons dissolve in QM.  However, this cannot be considered
as a probe of the transition because other mesons such as $u\bar d$ and
heavy $b \bar b$ dissolve at much higher density.  Thus, at least within this 
approach, there is no first order phase transition
and the transition from NM to QGP
is smooth.  The meson masses change depending on density and they
disappear at high enough densities due to the repulsive
parts of the potential and the Fermi motion which overcome the attractive
Coulombic plus chromomagnetic potentials.  The $\rho$ meson, for which
the c.t. is repulsive,  dissolves at all densities in QM.  
Experiments in p(or e,$\gamma, \pi$)
-Nucleus collisions, where we expect the density to be close at $\rho_0$ 
and T=0, can
give some further constraint on the physics involved.

% {\ \vskip 0.7 cm \centerline{\bf ACKNOWLEDGMENTS} }

%\end{multicols}

\newpage

{\ \centerline 
{\bf Table I} }

Quark masses used in this work.

\vspace{0.7cm}

\begin{tabular}{|c|c|}
\hline
$Quark Mass $ &GeV   \\ \hline
$u$ & 0.13  \\ 
$d$ & 0.13  \\ 
$s$ & 0.35   \\ 
$c$ & 1.45   \\
$b$ & 4.8   \\
$t$ & 180.    \\[3mm] \hline
\end{tabular}

\newpage
%\begin{figure}[tbp]
%\begin{center}
%mbox{{ \epsfysize=14truecm \epsfbox{jpg98_fig1.eps}}}
%\vskip 1.5cm
%\caption{Average density (left column)
%for a $\pi$(top), $n$(middle) and a $\Delta$
%(bottom) hadrons.  The average is over tp (squares) and
%over tp and time (full line). Similarly for the average potential (right 
%column).\label{fig:fig1}}
%\end{center}
%\end{figure}
\begin{figure}[tbp]
\begin{center}
\mbox{{ \epsfysize=14truecm \epsfbox{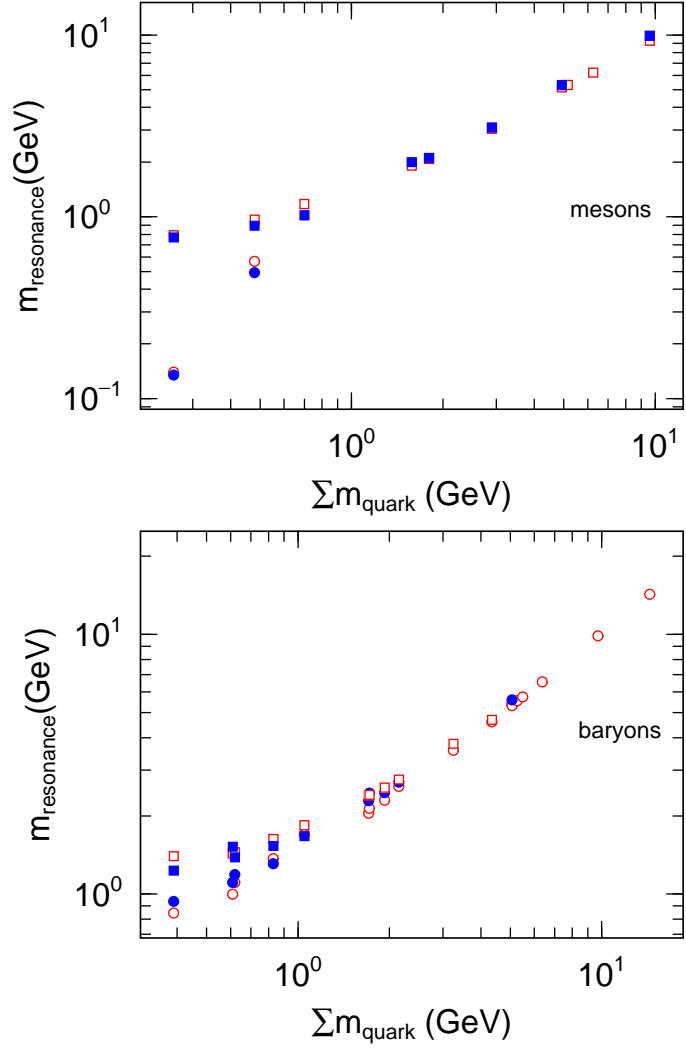}}}
\vskip 1.5cm
\caption{Meson masses (top) vs. masses of the $q\bar q$ pair. The symbols
refer to the observed (full symbols) and calculated (open
symbols) masses. The circles refer to the pseudoscalar and the
 squares to vector mesons masses. 
 Similarly for baryons (bottom). Data are taken from [12].\label{fig:fig1}}
\end{center}
\end{figure}
\begin{figure}[tbp]
\begin{center}
\mbox{{ \epsfysize=14 truecm \epsfbox{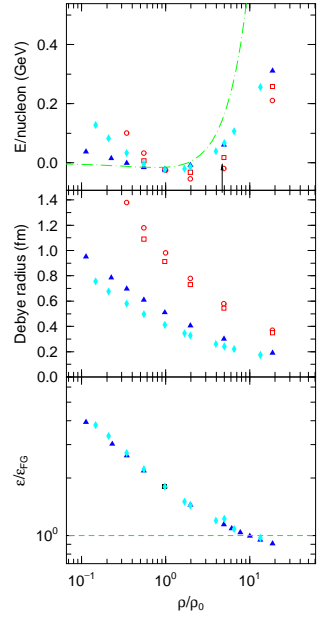}}}
\vskip 1.5cm
\caption{Energy per nucleon (top), Debye radius (middle) and energy density
of the quarks 
(bottom) vs. normalized density for various choices of the input parameters
(see text). The dashed-dotted curve (top) is obtained using a Skyrme
potential.\label{fig:fig2}}
\end{center}
\end{figure}
\begin{figure}[tbp]
\begin{center}
\mbox{{ \epsfysize=12 truecm \epsfbox{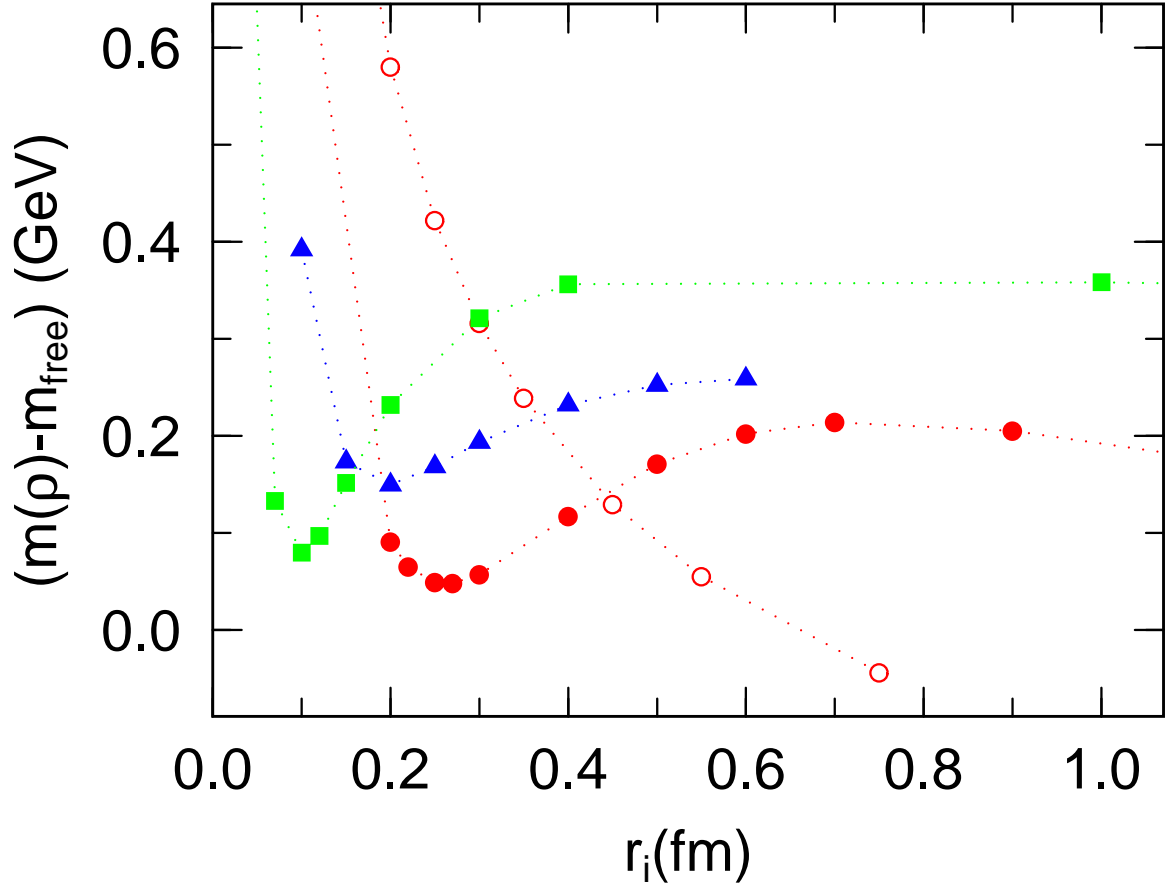}}}
\vskip 1.5cm
\caption{
Meson masses in medium (minus their free masses)
 vs the initial radius of the $q \bar q$
systems.  The open circles refer to the $\rho$ meson at $\rho=0.0235 fm^{-3}$,
full circles and triangles refer to the $\pi$ and $\eta$ at $\rho_0$,
the squares to $\gamma(1s)$ at $\rho=1.046 fm^{-3}$. \label{fig:fig3}}
\end{center}
\end{figure}
\begin{figure}[tbp]
\begin{center}
\mbox{{ \epsfysize=12 truecm \epsfbox{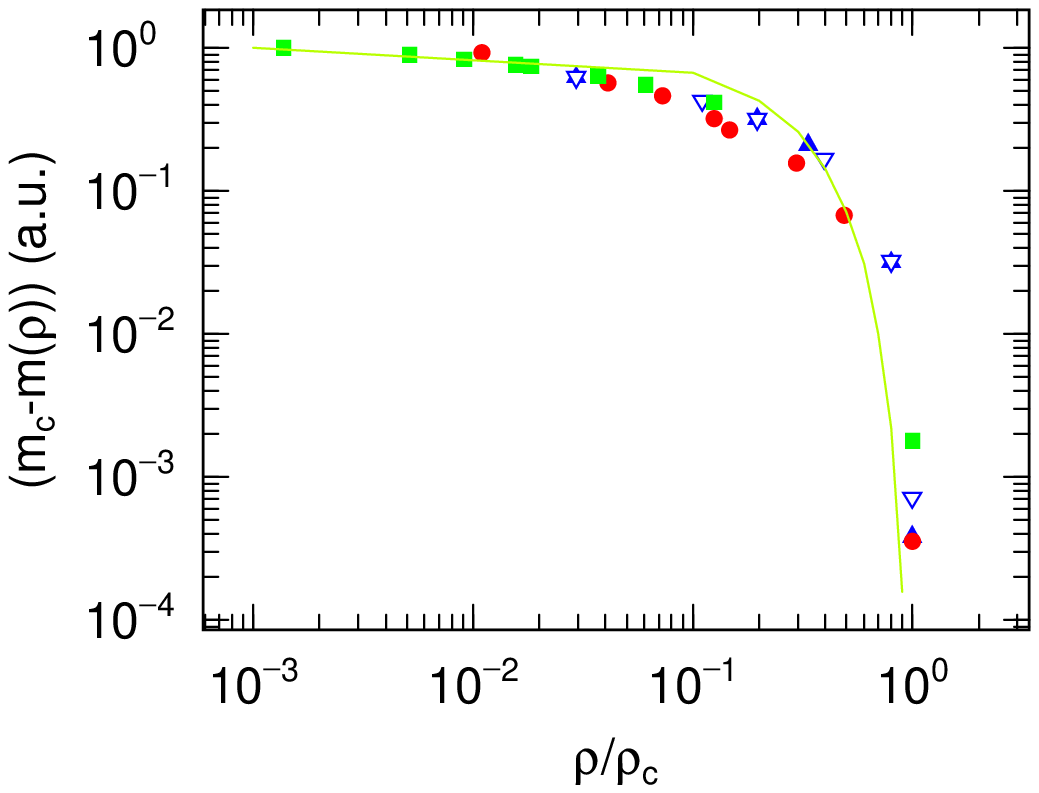}}}
\vskip 1.5cm
\caption{'Critical' minus meson mass calculated at $\rho$
vs. $\rho/\rho_c$ for $b\bar b$(squares),$c\bar c$ (triangles)
 and $u\bar d$ (circles).  The curve is a fit according to eq.(1).
\label{fig:fig4}}
\end{center}
\end{figure}

\begin{references}
\bibitem{nota}Such large values of $\alpha_0$ should not surprise.  
Infact recall that
a QCD result \cite{pov} gives $\alpha(Q^2) \propto 1/ln(Q^2/\Lambda^2)$.
 In QM the momentum transfer $Q\approx p_f\approx \Lambda$ at $\rho_0$, thus
we are in the region where the strong coupling constant can be very large.
\bibitem{nota0}For the sets
of parameters I and II we can easily fit the masses of the $\pi$ and the $n$
in vacuum by tuning opportunely the values of $\bar \alpha$ and $\sigma$ 
in eq.(3).    However,
set I is able to reproduce the masses of the $\rho$ and the
 $\Delta$ as well,see sect.III\cite{bon99},
 while set II gives masses 500-700 MeV larger than data for
those particles.  Heavier meson masses in vacuum are reproduced by fitting the
corresponding quark masses since the c.t. gives a vanishing contribution, 
eq.(3) \cite{bon99}.
\bibitem{nota1} For set II $\beta=3.8$, the ($m_c,\rho_c$) values are 
(9.79 GeV, 17.15  $fm^{-3}$),
(3.172 GeV,0.798 $fm^{-3}$), (3.174 GeV,0.8 $fm^{-3}$), 
(0.199 GeV, 2.14 $fm^{-3}$), for the $\gamma(1S)$, $\eta_c(1S)$, $J/\psi$ 
and the $\pi$ respectively. For set I the results are qualitatively the same.
\end{references}
\end{document}